\documentclass[10pt,aps,prl,twocolumn,showpacs,superscriptaddress,groupedaddress,floatfix,
nofootinbib]{revtex4-1}  

\usepackage{color}

\usepackage[active]{srcltx}

\usepackage{amsfonts}
\usepackage{amsmath}
\usepackage{amssymb}
\usepackage{latexsym}
\usepackage{graphicx}
\usepackage[english]{babel}

\newcommand{\HHH}{${\rm H}_3\,$}
\newcommand{\HHHp}{${\rm H}_3^+\,$}

\newcommand{\dd}{$^\dagger\,$}

\newcommand{\al}{\alpha}
\newcommand{\be}{\beta}

\newcommand{\Si}{\Sigma}

\newcommand{\De}{\Delta}

\newcommand{\rar}{\rightarrow}

\begin{document}

\title{ Finite Hydrogenic molecular chain \HHH \ and ion H${}_2^-$ exist
in a strong magnetic field}

\author{D.J.~Nader}
 \email{daniel.nader@correo.nucleares.unam.mx}
 \affiliation{Instituto de Ciencias Nucleares, Universidad Nacional Aut\'onoma de M\'exico,
 A. Postal 70-543 C. P. 04510, Ciudad de M\'exico, M\'exico}

 \author{J.C.~L\'opez~Vieyra}
 \email{vieyra@nucleares.unam.mx}
 \affiliation{Instituto de Ciencias Nucleares, Universidad Nacional Aut\'onoma de M\'exico,
 A. Postal 70-543 C. P. 04510, Ciudad de M\'exico, M\'exico}

\author{A.V.~Turbiner}
 \email{turbiner@nucleares.unam.mx}
 \affiliation{Instituto de Ciencias Nucleares, Universidad Nacional Aut\'onoma de M\'exico,
 A. Postal 70-543 C. P. 04510, Ciudad de M\'exico, M\'exico}

\begin{abstract}
The existence and stability of the linear hydrogenic chain \HHH \ and
H${}_2^-$ in a strong magnetic field is established. Variational
calculations for \HHH \ and H${}_2^-$ are carried out in magnetic
fields in the range $10^{11}\leq B \leq 10^{13}\,$G with 17-parametric
(13-parametric for H${}_2^-$), physically adequate trial function.
Protons are assumed infinitely massive, fixed along the
magnetic line.  States with total spin projection
$S_z=-3/2$ and magnetic quantum numbers $M=-3,-4,-5$ are studied. It
is shown that for both \HHH \ and H${}_2^-$ the lowest energy state corresponds
to $M=-3$ in the whole range of magnetic fields studied.  As for a magnetic field
$B \gtrsim  10^{11}\,$G both \HHH \ and H${}_2^-$  exist as metastable states, 
becoming stable for $B \geq 1.9 \times 10^{11}\,$G and for $B \geq 2.7 \times 10^{11}\,$G,
respectively. The excited states $^4(-4)^+$, $^4(-5)^+$ of ${\rm H}_3$
and H${}_2^-$ appear at magnetic fields $B > 7 \times 10^{11}$ and $10^{12}$\,G, 
respectively.
\end{abstract}

\maketitle

Surface magnetic fields $B\sim 10^{11}-10^{14}\,$G exist for
many neutron stars, while for magnetized white dwarfs the magnetic
field can reach $B \sim 10^{9}\,$G (see e.g. \cite{Science:2004,
Garcia:2016}).  Among these stars there are many which have
atmosphere, which is mostly composed from electrons and protons while
some heavy nucleus (e.g. $\al, O, Ne, C, Fe$) can be also present. It
is natural to assume that for surface temperatures of order of 10
{\it eV}, electrons and protons can condense into hydrogenic species,
forming Hydrogen atoms and Hydrogenic molecular ions.

These enormous magnetic fields modify dramatically the structure of
atoms and molecules: as the strength of the magnetic field increases
the atoms and molecules become more bound and more compact, their
electronic clouds get pronounced cigar-type form and eventually the
charged centers are aligned on a magnetic line. Thus, the Coulomb systems
become quasi-one-dimensional. Such strong magnetic fields eventually
lead to the appearance of exotic charged molecular and atomic systems
which do not exist without a strong magnetic field (see e.g.
\cite{Turbiner:2006} and references therein). In particular, the
pioneering studies by Ruderman \cite{Ruderman:1974} and,
independently, by Kadomtsev-Kudryavtsev
\cite{Kadomtsev:1970,Kadomtsev:1971a} predicted qualitatively that
finite and supposedly even infinite neutral hydrogenic (linear) chains
(and Wigner crystals) could exist if magnetic fields are sufficiently
strong. In particular, accurate calculations have shown that in
addition to the hydrogen atom ${\rm H}$ and the ${\rm H}_2^+$
molecular ion, which exist for any magnetic field, at magnetic fields
$B \gtrsim 10^{11}\,$G three protons situated along the magnetic line
can be bound by single electron (!) forming the exotic molecular ion
${\rm H}_3^{2+}$ in linear configuration
\cite{Turbiner:1999}. Furthermore, ${\rm H}_3^{2+}$ turns out to be
the most bound one-electron hydrogenic system for magnetic fields $B
\gtrsim 10^{13}\,$G \cite{Turbiner:2006}.  This discovery was used
later to construct a model of the atmosphere of the isolated neutron
star 1E1207.4-5209 (see ~\cite{Turbiner:2004m}) to explain the origin
of the absorption lines at $\sim 0.7$ and $1.4\,$KeV detected by
Chandra ~\cite{Sanwal:2002} and confirmed by
XMM-Newton~\cite{Bignami:2003} X-ray observatories. It predicts the
surface magnetic field of $(4 \pm 2)\times 10^{14}$\,G.

For systems with two or more electrons it is known that in
sufficiently strong magnetic fields the ground state appears in the
configuration where all spins of electrons are antiparallel to the
magnetic field direction, hence, the total spin takes its maximal
value. In field free case the total spin usually does not take the
maximal value. It implies that the ground state type can evolve with
magnetic field strength. This phenomenon was quantitatively observed
for the first time for ${\rm H}_2$ molecule, where it was shown that
the ground state changes from spin-singlet state ${}^1\Si_g$ at zero
and weak magnetic fields to triplet unbound (repulsive) state
${}^3\Si_u$ for intermediate fields $B\gtrsim 5\times 10^8\,$G\,\,
while for larger magnetic fields $B \gtrsim 3\times 10^{10}\,$G the
ground state is the spin-triplet state ${}^3\Pi_u$~\cite{Detmer:1997}
(and references therein). It implies that in  this domain ${\rm H}_2$
molecule is unstable towards dissociation ${\rm H} + {\rm H}$.
As for the molecular ion \HHHp in a strong magnetic field in linear
parallel configuration (aligned with the magnetic field direction)
the ground state of \HHHp changes from spin-singlet ${}^1\Si_g$ state
for weak magnetic fields $B\lesssim 5\times 10^{8}\,$G (at
magnetic fields close to zero the ground state of \HHHp is equilateral
triangular configuration in spin-singlet state, it is the most bound
hydrogenic specie) to a weakly-bound spin-triplet
${}^3\Si_u$ for intermediate fields and, eventually, to spin-triplet
state ${}^3\Pi_u$ for magnetic fields $B\gtrsim 10^{10}\,$G,
it is always stable \cite{Turbiner:2007}.
The list of \hbox{one-,} two-electron hydrogenic systems, which can exist in a
strong magnetic field being stable, is given in
\cite{Turbiner:2006, Turbiner:2010}.

In general, it is known very little about atomic-molecular systems
with three electrons in a strong magnetic field. In particular, the
neutral hydrogenic chain \HHH was studied in ~\cite{Lai-Salpeter:1992}
along with finite hydrogenic chains ${\rm H}_n\,,\, \scriptstyle
n=2,3,4 \ldots$ in strong magnetic fields $B \geq 10^{11}$\,G in
sophisticated multiconfigurational Hartree-Fock method. The accuracy
obtained was limited, grossly overestimated and in some cases the
results were indicated as non-reliable (spurious), the question of the
existence and stability was never discussed. We are not aware on any
studies of negative hydrogenic molecular ion H$_2^-$.

\begin{figure}
\begin{center}
\includegraphics[angle=0,width=60mm]{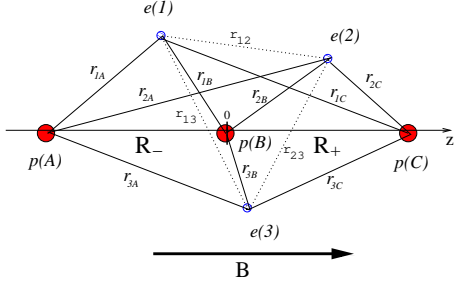}
 \caption{\label{H3inline} Geometrical settings and notations for the linear chain
 $\rm{H}_{3}$ in magnetic field $\bf{B}$ parallel to the molecular axis.
}
\end{center}
\end{figure}

In present Letter the linear molecular chain \HHH$\,$ and the H$_2^-$ ion
are studied in a strong magnetic field being situated along a magnetic
line (we call it the {\it parallel configuration}, see Fig. \ref{H3inline}).
Magnetic field is assumed strong enough to have minimal total spin projection
$S_z=-3/2$, thus, we focus on states with total spin 3/2, and total
magnetic quantum numbers $M=-2,-3,-4, -5$.  We explore the magnetic fields
$10^{11}\,\rm G \leq B\leq 10^{13}\,$G, where a non-relativistic
approach with static nucleus is still valid, see for discussion
\cite{Lai-Salpeter:1992}. The study is developed in the
Born-Oppenheimer approximation of zero order, i.e. the nuclei are
considered to be infinitely massive. The main goal of the study is to
localize the domain of existence and stability of \HHH and H$_2^-$ in strong
magnetic fields.
We consider two main dissociation channels for both \HHH:
(i) ${\rm H}_3\rar {\rm H}_2+{\rm H}$,
(ii) ${\rm H}_3 \rar {\rm H}_3^+ + {\rm e}$, and H$_2^-$:
(iii) ${\rm H}_2^-\rar {\rm H}^- + {\rm H}$,
(iv) ${\rm H}_2^- \rar {\rm H}_2 + {\rm e}$\,. Dissociation
energies are defined accordingly,
$E_{diss}^{(i)}=E_{T}^{{\rm H}_2+{\rm H}}-E_{T}^{{\rm H}_3}\,,$
$E_{diss}^{(ii)}=E_{T}^{{\rm H}_3^+ + e}-E_{T}^{{\rm H}_3}\,$, \ and
$E_{diss}^{(iii)} = E_{T}^{{\rm H}^- + {\rm H}} - E_{T}^{{\rm H}_2^-}\,,$
$E_{diss}^{(iv)} = E_{T}^{{\rm H}_2 + e} - E_{T}^{{\rm H}_2^-}\,$.
Note that writing about dissociation we ignore the conservation
of total angular momentum projection due to possible presence of
photon(s) in the final state and assume each of the final products
is in the lowest energy state. The energy of a free electron in a
magnetic field with $S_z=-1/2$ is zero (see \cite{He2pTurbinerVieyra}
for details).
The lowest longitudinal vibrational states around the
equilibrium configuration of the ground state are briefly
studied.

Atomic units are used throughout ($\hbar = m_e = e = 1$) in making calculations.
For the magnetic field $B$ written in a.u. the conversion $1\,
\mbox{a.u.}\, =2.35 \times 10^9$\,G is used, as for the energy $1\, a.u. =
27.2$\,eV\,.

\noindent
{\it Generalities.}\ The Hamiltonian describing the Coulomb system of
three electrons and three aligned (infinitely) massive charged centers
$\tt A,B,C$ subject to a constant uniform magnetic field, parallel to
the molecular axis \hbox{${\bf B} = B\,\hat{e}_{z}$}, is given by
\begingroup
\small
\begin{align}
 \label{Ham}
 \mathcal{H} &= -\sum_{i=1}^{3}\left(\frac{1}{2}\De_{i} +
 \sum_{\scriptstyle \eta={\tt A,B,C}}\frac{Z_{\eta}}{r_{i\eta}}\right) +
 \sum_{i=1}^{3}\sum_{j>i}^{3}\frac{1}{r_{ij}} +
 \frac{B^2}{8}\sum_{i=1}^{3}\rho_{i}^2
 \nonumber \\ &
 + \frac{B}{2}(L_{z} + 2S_{z}) + \frac{Z_{\tt A}Z_{\tt B}}{R_{-}}
 + \frac{Z_{\tt A}Z_{\tt C}}{R_{+}+R_{-}}
 + \frac{Z_{\tt B}Z_{\tt C}}{R_{+}}\,,
\end{align}
\endgroup
where $\De_{i}$ is Laplacian for the $i$-th electron, $i=1,2,3$.
$Z_{\tt A,B,C}=1$ are the charges of the heavy centers, $\eta = \tt A,
B, C$ (protons).  The term $-\frac{Z_{\eta}}{r_{i\eta}}$ corresponds
to the Coulomb attractive potential between the electron $i$ and the
nuclei $\eta$ with $r_{i\eta}$ being the electron-nuclei distance. The
term $\frac{1}{r_{ij}}$ stands for the inter-electron repulsion
between electrons $i$ and $j$, where $r_{ij}$ is the distance between
electrons.  In turn, $\frac{Z_{\tt A}Z_{\tt B}}{R_{+}}$,
$\frac{Z_{\tt A}Z_{\tt C}}{R_{+}+R_{-}}$ and $\frac{Z_{\tt B}Z_{\tt C}}{R_{-}}$
are the classical Coulomb repulsion energy terms between the (fixed)
$\tt A, B, C$ nuclei. The internuclear (classical) distances are $R_+$
and $R_-$ (see Fig.\ref{H3inline}). The Hamiltonian (\ref{Ham})
includes the Zeeman terms $\frac{1}{2}{\bf B}\cdot{\bf L}=
\frac{1}{2}B\,L_z$, and $\frac{g}{2}{\bf B}\cdot{\bf S} = B\,S_z$
(with the g-factor, $g=2$) and the diamagnetic terms
$\frac{B^2}{8}\rho_{i}^2\,,$ where $\rho_{i}^2=x_{i}^2+y_{i}^2\,, (i=
1,2,3)$ is the transverse distance (squared) between the $i$-th
electron and the molecular axis.
Putting $Z_{\tt B}=0$ the Hamiltonian (\ref{Ham}) describes the
3-electron 2-center system in a magnetic field which may correspond to H$_2^-$ molecule.

For the linear ${\rm H}_3$ molecule the equilibrium configuration
was assumed symmetric: $R_{+}=R_{-}
\equiv R$ \hbox{-- it} is confirmed by calculations. In this configuration the
Hamiltonian (\ref{Ham}) is invariant under transformations $z_{1,2,3}
\rar -z_{1,2,3}$, and also invariant under rotation around the
molecular axis. Thus, the conserved quantities we take into account
are: the parity ($\pm$) w.r.t. to $z \rar -z$, see Fig.1, the $z$-component of
the total angular momentum $M_z$ (magnetic quantum number), the total
electronic spin $S$ and its projection $S_z$ along the $z$-axis. These
conserved quantities characterize the state of the system in the
standard spectroscopic notation $^{2S+1}(M_z)^{\pm}$.

The variational method is applied to find the energy of some low-lying
states using a trial function based on physics relevance
criteria. These trial functions allow us to reproduce both the Coulomb
singularities and the correct asymptotic behavior of the potential at
large distances (see, e.g. \cite{Turbiner:1984}).

\noindent
{\it Trial Function.}\ The orbital (spatial) part of the trial
function is proposed as a product of screened lowest Coulomb
orbitals, Landau orbitals and exponential correlation factors:
\begin{align}
 & \psi({\bf r}_{1},{\bf r}_{2},{\bf r}_{3}) =
 \, e^{\al_{12}r_{12}+\al_{13}r_{13}+\al_{23}r_{23}}\, \label{function} \\
 & \qquad \times  \left( \prod_{k=1}^{3} \rho_{k}^{|m_{k}|}
 e^{im_{k}\phi_{k}}\,
 e^{-\al_{k\tt A}r_{k\tt A}-\al_{k\tt B}r_{k\tt B}-
 \al_{k\tt C}r_{k\tt C}-\frac{B}{4}\be_{k}\rho_{k}^2}\right)\,,
 \nonumber
\end{align}
where $\rho_{k}$, $\phi_{k}$ are the radial/angular cylindrical
coordinates of the $k$-th electron with the magnetic quantum number
$m_k$, $k=1,2,3$.
In turn, the parameters $\al_{k\eta}$, $\al_{ij}$ with $i<j=1,2,3$,
$\eta=\tt A,B,C$ are screened effective charges, $\be_{1,2,3}$ measure
screening of the magnetic field, the classical internuclear distances
$R_{\pm}$ are taken as variational parameters. The total number of
variational parameters in (\ref{function}) is $17$ (13 as
for ${\rm H}_2^-$).

For states of total spin $S=3/2$ with projection $S_z=-3/2$, the spin
part of the trial function is \( { \chi= \beta(1)\beta(2)\beta(3)}\ ,
\) where ${\beta(k)}\,,\ k=1,2,3$ represents the $S=1/2$ spinor of the
$k$-th electron with negative spin projection. Thus, the properly
symmetrized total wave function is given by
\begin{equation}
\label{totalfunction}
 \Psi({\bf r}_{1},{\bf r}_{2},{\bf r}_{3})=(1+\sigma_{N}\hat{P}_{\tt AC})
 \hat{A}[\psi({\bf r}_{1},{\bf r}_{2},{\bf r}_{3})\chi]\,,
\end{equation}
where $\hat{P}_{\tt AC}$ is the operator of permutation of the two end
nucleus $\tt A$ and $\tt C$ (see Fig.\ref{H3inline}), and
$\sigma_{N}=\pm1$ is $z$-parity.  The operator $\hat{A}$ is the three
particle antisymmetrizer
$$\hat{A}=1-\hat{P}_{12}-\hat{P}_{13}-\hat{P}_{23}+\hat{P}_{231}+\hat{P}_{312}\,,$$
acting on the coordinates of the electrons. Here $\hat{P}_{ij}$ is
the operator of permutation of the electrons $i\leftrightarrow j$,
and $\hat{P}_{ijk}$ stands for the permutation of $(123)$ into $(ijk)$.

Numerics for the calculation of the variational energy is described in
\cite{He2pTurbinerVieyra}. Computations were performed in parallel
using the cluster Karen (ICN-UNAM) with 120 Intel Xeon processors at
$\sim 2.70\,$GHz.

\bigskip
\noindent
{\it Results.}\ Variational calculations of the system $(3p,3e)$
\ and $(2p,3e)$ in parallel configuration, see Fig.1, for magnetic fields
$10^{11}\,$G$\leq B\leq 10^{13}\,$G are carried out using the
trial function (\ref{function}) for spin-quartet states with total
spin 3/2 and for total magnetic quantum numbers $M=-2,-3,-4, -5$.
For $M=-2$ both systems are unbound: one electron always goes to infinity in $z$-direction.
In all other cases we find bound states with a well pronounced minimum in the energy
at finite nuclear distances at $R_+=R_-$ for \HHH and for H$_2^-$ systems.
It is seen that the total energy and the equilibrium internuclear distances
of \HHH \ and H$_2^-$\ decrease monotonously with the magnetic field increase,
the systems become more bound and compact, see Table \ref{tableH3} and \ref{tableH2m},
respectively. Among the states we study the state with $M=-3$ when the magnetic quantum
numbers are $m_1=0,m_2=-1,m_3=-2$ (being in agreement to the Pauli
principle, it guarantees zero Pauli force) realizes the {\it ground state} of
both \HHH and ${\rm H}_2^-$.  Taking different combinations of $m_i$ keeping
$M=-3$ in (\ref{function}) does not improve the energies. States with
$M=-4,-5$ always lie above the ground state $^4(-3)^+$. Note that for this state
the equilibrium distance is smallest with respect to other states, 
see Table \ref{tableH3} and \ref{tableH2m}.
Dissociation energies as well as lowest longitudinal vibrational energies
are presented in Table \ref{tableH3} and Table \ref{tableH2m}.
Systematically, the total energies for \HHH \ are smaller than ones 
calculated in \cite{Lai-Salpeter:1992}, while equilibrium distances are comparable.

For $B \gtrsim 10^{11}$\,G the molecule \HHH is always stable towards
the dissociation channel (i) ${\rm H}_3^+ + e$, see Table \ref{tableH3}.
However, as for the channel (ii) ${\rm H}_3 \rar {\rm H}_2+{\rm H}$ there
is a critical magnetic field, where dissociation energy for the ground 
state vanishes,
\begin{equation}
\label{BcH3}
    B_c^{{\rm H}_3}(0) \sim 1.9\times 10^{11}\mbox{G}\,,
\end{equation}
It indicates that the linear molecular chain \HHH always exists,
it becomes {\it stable} at $B > B_c^{{\rm H}_3}(0) $ being {\it metastable}
for smaller magnetic fields. The stability of the linear molecule \HHH\
was checked towards longitudinal symmetric vibrations (s), $R_+=R_-$ 
and antisymmetric vibrations
(a) $R_+ \neq R_-$. The lowest vibrational energy of the symmetric mode
$E^{(s)}_{vib}$ are always smaller than those of the antisymmetric mode
$E^{(a)}_{vib}$. Both energies increase with the magnetic field growth,
see Table \ref{tableH3}. For $B \gtrsim 5 \times 10^{11}\,$G the
dissociation energies are always larger than the sum of the lowest
vibrational energies, hence, the potential energy surface contains,
at least, one vibrational state.

As for ${\rm H}_2^-$ ion the total energies, equilibrium distances and
dissociation energies as well as lowest longitudinal vibrational energy
are presented in Table \ref{tableH2m}.
For $B \sim 10^{11}$\,G\ ${\rm H}_2^-$ is unstable towards both channels
(iii) and (iv). However, at the magnetic field $B \sim 2.35\times 10^{11}\,$G
the channel (iii) gets forbidden while the channel (iv) is still open.
With magnetic field increase at
\begin{equation}
\label{BcH2-}
B_c^{{\rm H}_2^-}(0) \sim 2.7 \times 10^{11}\mbox{G}\,,
\end{equation}
this channel gets also forbidden and H$_2^-$ becomes {\it stable},
being {\it metastable} for smaller magnetic fields.
The lowest vibrational energy increases with the magnetic field growth.
Potential energy surface contains, at least, one vibrational state in all range
of magnetic fields studied.

\medskip
\noindent
{\it Excited States.}\  The excited states $^4(-4)^+$, $^4(-5)^+$
were studied for both \HHH and H$_2^-$. For $B \gtrsim  10^{11}\,$G
both states are bound for both systems.
It was checked that minimal energy for \HHH \ always corresponds
to symmetric configuration $R_+=R_-$.
The total energy and the equilibrium internuclear
distance $R_{eq}$ for \HHH and H$_2^-$ are presented in
Table~\ref{table3} and ~\ref{table4}, respectively.
For both states $^4(-4)^+$, $^4(-5)^+$ for each system there exist
critical magnetic fields, see below section {\it Conclusions},
for which they become stable with respect to dissociation
${\rm H}_3 \rightarrow {\rm H}_2+{\rm H}$ and
${\rm H}_2^- \rightarrow {\rm H}_2+e$, correspondingly.
\begin{table}[htb]
\resizebox{0.9\columnwidth}{!}{%
\begin{tabular}{ccccccc} \hline \hline
$B\,(10^9\,{\rm G})$  & $\, E_{T}^{{\rm H}_3}$
       & $\, R_{eq}^{{\rm H}_3}$
       & $\, E_{vib}^{0(s)}$
       & $\, E_{vib}^{0(a)}\,$
       & $\quad E_{diss}^{{\rm H}_3  \rar {\rm H}_2+{\rm H}}$
       & $E_{diss}^{{\rm H}_3  \rar {\rm H}_3^+ + e}$
       \\[3pt]
       \hline \hline
$100.0$               &    -8.62         &    0.482      &           &             &      -           & -      \\
                      &    -8.50\dd      &    0.48       &           &             &                  &        \\
$117.5 $              &    -9.18         &    0.460      &           &              &       -0.174     & -      \\
$235.0$               &    -12.09        &    0.359      &  0.073    &   0.150      &        0.127     & 2.63   \\
$500.0$               &    -16.15        &    0.277      &  0.104    &   0.208      &        0.644     & -      \\
                      &    -16.08\dd     &    0.27\dd    &           &              &                  &        \\
$1000.0$              &    -20.90        &    0.219      &  0.154    &   0.282      &        1.382     & -      \\
                      &    -20.81\dd     &    0.22\dd    &           &              &                  &        \\
$2350.0$              &    -28.39        &    0.168      &  0.237    &   0.425      &        2.759     & 6.12   \\
$5000.0$              &    -36.78        &    0.134      &  0.337    &   0.601      &        -         & -      \\
                      &    -36.77\dd     &    0.13\dd    &           &              &                  &        \\
$10000.0$             &    -46.19        &    0.110      &  0.467    &   0.814      &        -         & -      \\
                      &    -46.19\dd     &    0.11\dd    &           &              &                  &        \\
\hline
\end{tabular}
}
 \caption{
   \label{tableH3}
    Ground state of ${\rm H}_3$: total energy $E_T$ and equilibrium distance $R_{eq}$
    in symmetric configuration $(R_+=R_-)$ {\it vs} magnetic field for 
    the ground state $^4(-3)^+$.
    Results \cite{Lai-Salpeter:1992} marked by$\dagger$ as for results for H$_2$ \cite{Turbiner:2010}, H$_3^+$ \cite{Turbiner:2007} and H \cite{Kravchenko:1996}.
   Lowest longitudinal vibrational energies for (anti)symmetrical $(s)$ and
   $(a)$ modes presented as well as dissociation energies into
   two decay channels. Energies and distances in a.u.
   }
\end{table}

\begin{table}[htb]
\begin{tabular}{cccccc} \hline \hline
$B\,(10^9\,{\rm G})$
       & $E_{T}^{{\rm H}_2^-}$
       & $\, R_{eq}^{{\rm H}_2^-}$
       & $\, E^0_{vib}$
       & $E_{diss}^{{\rm H}_2^-\rar  {\rm H}^- + {\rm H}}$
       & $E_{diss}^{{\rm H}_2^-\rar {\rm H}_2+e}$
       \\[3pt]
       \hline \hline
$235.0$            &   -8.02  & 0.352 &\        & 0.05   & -0.16 \\
$500.0$            &   -11.18 & 0.300 &\ 0.11   & -      & 0.49  \\
$1000.0$           &   -14.29 & 0.247 &\ 0.16   & -      & 0.74  \\
$2350.0$           &   -18.92 & 0.188 &\ 0.26   & 2.79  & 0.95  \\
$5000.0$           &   -23.85 & 0.151 &\ 0.37   & -      & -      \\
$10000.0$          &   -29.26 & 0.128 &\ 0.52   & -      & -      \\
\hline
\end{tabular}
 \caption{
   \label{tableH2m}
  Ground state ${\rm H}_2^-$: total energy $E_T$ and equilibrium distance $R_{eq}$
   for the state $^4(-3)^+$ (ground state) of ${\rm H}_2^{-}$.
   Lowest longitudinal vibrational energy presented as well as dissociation energies into
   two decay channels. Energies and distances in a.u.}
\end{table}
\begin{table}[htb]
{%
\begin{tabular}{cccccccc} \hline \hline
$B\, (10^9\,{\rm G})$
                       & $E_{T}^{{\rm H}_3 [\scriptstyle {}^4(-4)^{+}]}$
                       & $\, R_{eq}^{{\rm H}_3 [\scriptstyle {}^4(-4)^{+}]}$
                       & \phantom{--}
                       & $E_{T}^{{\rm H}_3 [\scriptstyle {}^4(-5)^{+}]}$
                       & $\, R_{eq}^{{\rm H}_3 [\scriptstyle {}^4(-5)^{+}]}$
                       \\[2pt]
       \hline \hline
$100$                 &   -8.15    &    0.50  &&        &      \\
$500$                 &  -15.31    &    0.30  &&        &      \\
$1000$                &  -19.89    &    0.23  && -19.37 & 0.24 \\
$2350$                &  -26.82    &    0.18  && -26.29 & 0.19 \\
\hline
\end{tabular}
}
 \caption{
   \label{table3}
   ${\rm H}_3$ (excited states): Total energy $E_T$ and equilibrium distance $R_{eq}$ 
   (for $R_+=R_-$) for states $^4(-4)^+$ and $^4(-5)^+$. Energies and distances in a.u.
   }
\end{table}
%

\noindent
{\it Conclusions.\,} We have shown that trihydrogen molecule \HHH in the form of
linear chain exists for $B \gtrsim  10^{11}\,$G with the ground state $^4(-3)^+$.
It becomes stable towards dissociation for magnetic fields  larger than the critical
magnetic field $B_c^{\rm H_3}(0) \simeq 1.9 \times 10^{11}\,$G\,,
it also remains stable towards small longitudinal vibrations.
Its first excited state $^4(-4)^+$ is always bound and appears stable towards dissociation
at $B_c^{\rm H_3}(1)\sim 6.9\times 10^{11}\,$G, while the second excited state $^4(-5)^+$
is stable toward dissociation for $B \gtrsim B_c^{\rm H_3}(2) \sim 1.2\times 10^{12}\,$G.
It excludes a qualitative prediction \cite{Ruderman:1974} about existence of infinite
hydrogenic chain (and Wigner crystal) for magnetic fields $B < B_c^{\rm H_3}(0)$
and, in fact, $\lesssim 10^{12}$\,G.
For any available strong magnetic field the absorption due dissociation of \HHH occurs
at less than 100 eV, hence, could not be detected by Chandra or XMM X-ray observatories.
As for linear molecular chain H$_4$ one can estimate following the results of \cite{Turbiner:2010} for H$_2$, Table \ref{tableH3} for H$_3$ and \cite{Lai-Salpeter:1992} 
for H$_4$ that it becomes stable for $B \gtrsim 10^{12}$\,G.
Dihydrogen negative molecular ion H$_2^-$ with the ground state $^4(-3)^+$ is always bound
and becomes stable towards dissociation for $B_c^{\rm H_2^-}(0) \simeq 2.7 \times 10^{11}\,$G.
Its first excited state $^4(-4)^+$ as well as the second excited state $^4(-5)^+$ are metastable
towards dissociation, both states become stable at $B_c^{\rm H_2^-}(1)\sim 9.0\times 10^{11}\,$G
and $B \gtrsim B_c^{\rm H_2^-}(2) \sim 1.9\times 10^{12}\,$G, respectively.
Both molecular systems can be present in the magnetized neutron star atmosphere 
at large surface magnetic field but with not very hot surface temperature.

\begin{table}[hbt]
{%
\begin{tabular}{cccccc} \hline \hline
$B\, (10^9\,{\rm G})$
                       & $E_{T}^{{\rm H}_2^- [\scriptstyle {}^4(-4)^{+}]}$
                       & $\, R_{eq}^{{\rm H}_2^- [\scriptstyle {}^4(-4)^{+}]}$
                       & \phantom{--}
                       & $E_{T}^{{\rm H}_2^- [\scriptstyle {}^4(-5)^{+}]}$
                       & $\, R_{eq}^{{\rm H}_2^- [\scriptstyle {}^4(-5)^{+}]}$
                       \\[2pt]
       \hline \hline
$352.5$                       &  -9.34  &  0.33    &&\         &        \\
$500$                         & -10.72  &  0.30    &&\         &        \\
$1000$                        & -13.72  &  0.26    &&\ -13.46 &  0.27  \\
$2000$                        & -17.30  &  0.21    &&\ -17.06 &  0.22 \\
$2350$                        & -18.22  &  0.20    &&\ -17.99 &  0.20 \\
\hline
\end{tabular}
}
 \caption{
   \label{table4}
 ${\rm H}_2^-$ (excited states): Total energy $E_T$ and equilibrium distance $R_{eq}$
   for states $^4(-4)^+$, $^4(-5)^+$. Energies and distances in a.u.
   }
\end{table}
\noindent
{\it Acknowledgements.\,} The authors thank CONACyT
A1-S-17364 and DGAPA IN108815 grants (Mexico) for
partial support.
\bibliographystyle{apsrev4-1}

\end{document}